\documentclass[aps,twocolumn,showpacs,preprintnumbers,amsmath,amssymb,superscriptaddress]{revtex4}

\usepackage{amsmath}%
\usepackage{amssymb}%
\usepackage{dsfont}
\usepackage{graphicx}
\usepackage{dcolumn}
\usepackage{bm}

\providecommand{\msii}{\mbox{m/s$^2$}\,}
\providecommand{\refkl}[1]{(\ref{#1})}
\providecommand{\erw}[1]{\mbox{$\langle #1 \rangle$}}
\providecommand{\abl}[2]{\frac{{\rm d} #1}{{\rm d} #2}}  
\providecommand{\ablpart}[2]{\frac{\partial #1}{\partial #2}}  
\providecommand{\ablparttwo}[2]{\frac{\partial^{2} #1}{\partial #2^{2}}}
\providecommand{\sub}[1]{_{\rm #1}}

\begin{document}

\title{Understanding widely scattered traffic flows, the capacity
drop, platoons, and times-to-collision as effects of variance-driven time gaps}
\author{Martin Treiber}
  \email{martin@mtreiber.de}
  \homepage{http://www.traffic-simulation.de}
  \affiliation{Dresden University of Technology, Andreas-Schubert-Str. 23, 01062 Dresden, Germany}
\author{Arne Kesting}
  \email{kesting@vwi.tu-dresden.de}
  \affiliation{Dresden University of Technology, Andreas-Schubert-Str. 23, 01062 Dresden, Germany}
\author{Dirk Helbing}
   \email{helbing1@vwi.tu-dresden.de}
   \homepage{http://www.helbing.org}
   \affiliation{Dresden University of Technology, Andreas-Schubert-Str. 23, 01062 Dresden, Germany}
   \affiliation{Collegium Budapest -- Institute for Advanced Study,
   Szenth\'aroms\'ag u. 2, H-1014 Budapest, Hungary}

\date{\today}
\begin{abstract}
We investigate the adaptation of the time headways in car-following
models as a function of the local velocity variance, which is a
measure of the inhomogeneity of traffic flow.  We apply this mechanism
to several car-following models and simulate traffic breakdowns in
open systems with an on-ramp as bottleneck.  Single-vehicle data
generated by several 'virtual detectors' show a semi-quantitative
agreement with microscopic data from the Dutch freeway A9. This
includes the observed distributions of the net time headways and
times-to-collision for free and congested traffic. While the
times-to-collision show a nearly universal distribution in free and
congested traffic, the modal value of the time headway distribution is
shifted by a factor of about two in congested conditions.
Macroscopically, this corresponds to the 'capacity drop' at the
transition from free to congested traffic. Finally, we explain the
wide scattering of one-minute flow-density data by a self-organized
variance-driven process that leads to the spontaneous formation and
decay of long-lived platoons even for deterministic dynamics on a
single lane.
\end{abstract}

\maketitle

\section{\label{sec:intro}Introduction}

\begin{figure}
  \begin{center}
     \includegraphics[width=88mm]{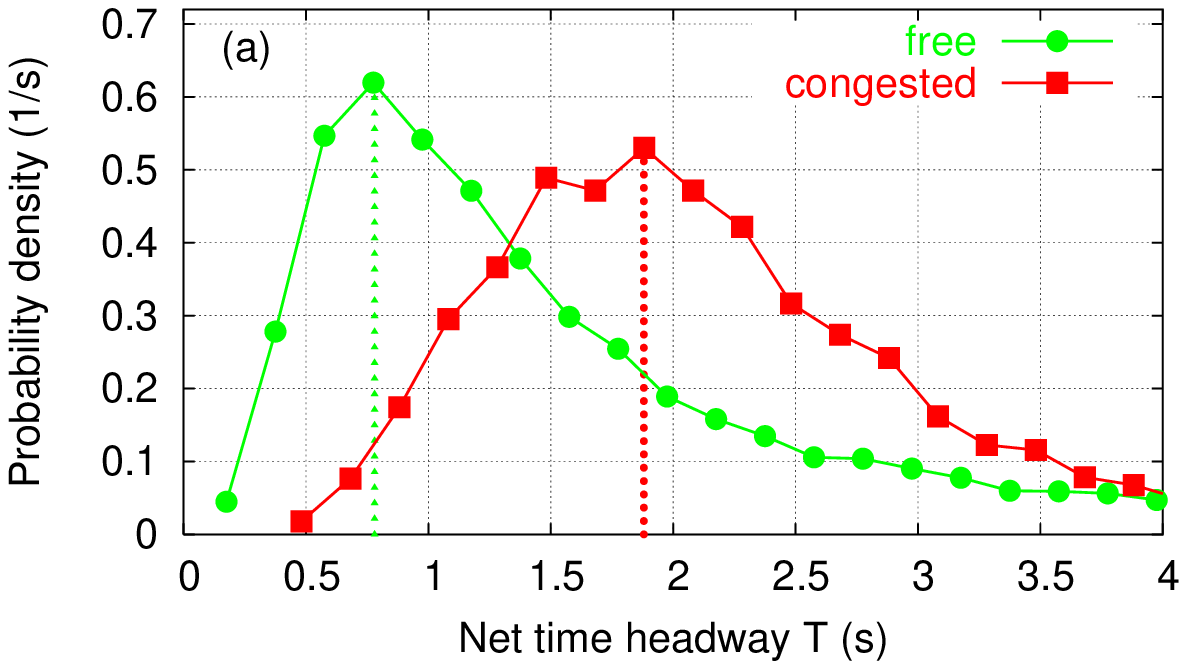} 
     \includegraphics[width=88mm]{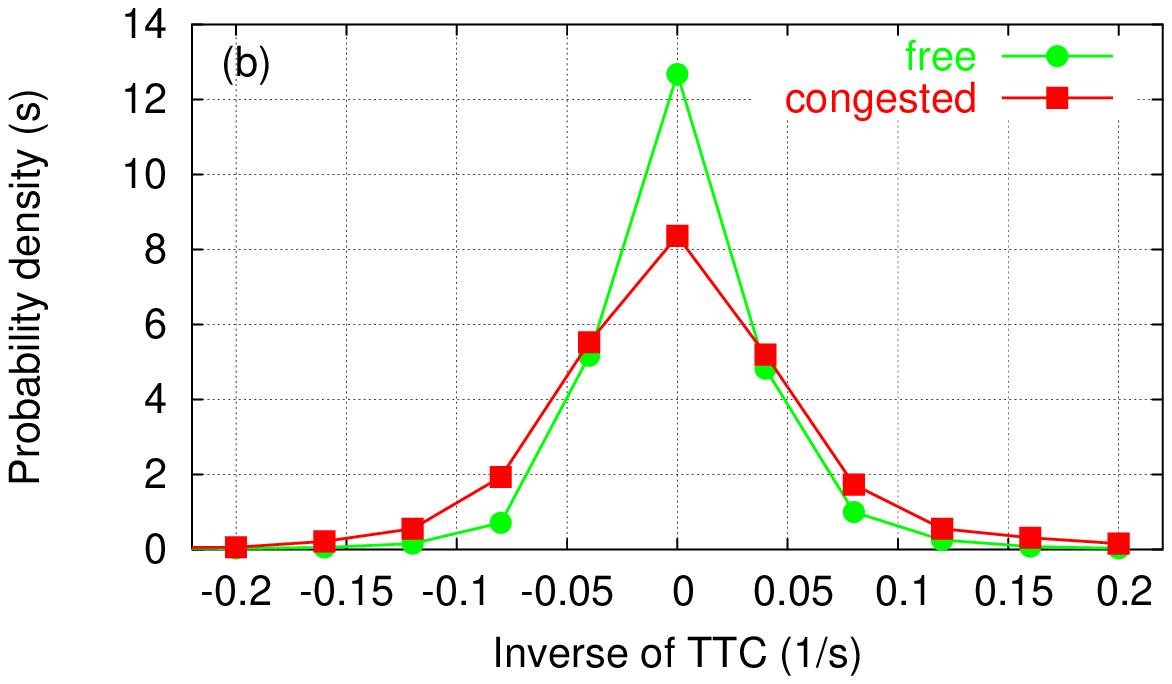}
     \includegraphics[width=88mm]{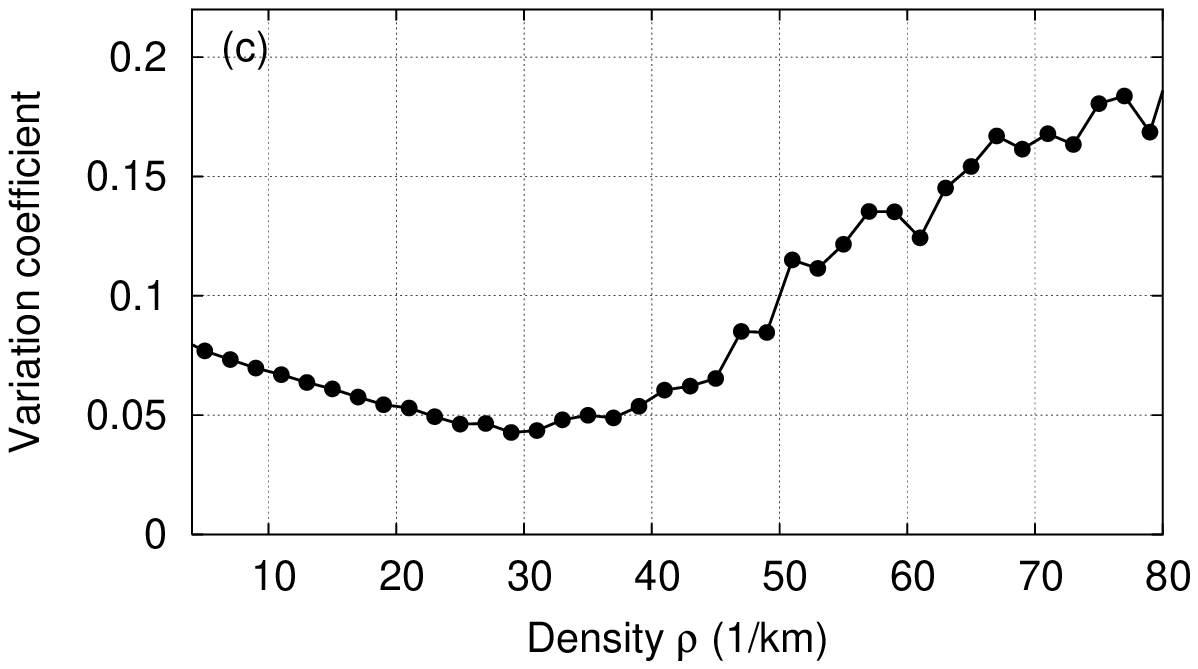}
  \end{center}

 \caption{\label{figEmp} Empirical statistical properties of cars
 following any kind of vehicle obtained from single-vehicle data from
 the left lane of the Dutch freeway A9 at a detector cross section 1.0
 km upstream of an on-ramp. (a) Net time headway according to
 Eq. \protect\refkl{Talpha}; (b) Inverse times-to-collision according
 to Eq. \protect\refkl{TTC}; (c) Variance coefficient $V_n$ according
 to Eq. \protect\refkl{Varcoeff}, as a function of the density.  In
 (a) and (b), the data set for 'free traffic' includes all
 single-vehicle data where the one-minute average of velocities was
 above 20 m/s, and the traffic flow above 1000 vehicles/h. 'Congested
 traffic' includes all data where the one-minute average of the
 velocities was below 15 m/s.  }

\end{figure}

One of the open questions of traffic dynamics is a microscopic
understanding of the observed wide variations in the time-headway
distributions \cite{Tilch-TGF99,Neubert-emp-99} that are closely
related to the wide scattering of flow-density data in the congested
regime \cite{Kerner-Rehb96-2,Katsu03}, see, e.g.,
Refs.~\cite{Kerner-book,Helb-opus} for an overview. Apart from their
wide variations, the average values of the time headways depend
strongly on the traffic density. For congested traffic, the modal
value, i.e., the value where the distribution has its maximum, is
larger by a factor of about 2 compared to free traffic.  Figure
\ref{figEmp}(a) shows a typical example obtained from single-vehicle
detector data of the Dutch freeway A9 from Haarlem to Amsterdam.

With the increasing availability of single-vehicle data
\cite{Tilch-TGF99,Neubert-emp-99,Kno02-data}, 
further statistical properties of traffic became the subject of
investigation such as the velocity variance as a function of the
traffic density \cite{GKT}, or the distribution of the times-to-collision
(TTC), which plays an important role for traffic safety
\cite{Minderhoud-TTC,Hirst}.

In this paper, we therefore propose a variance-driven adaptation
mechanism, according to which drivers increase their safety time gaps
$T$ when the local traffic dynamics is unstable or largely
varying. This adaptation is, e.g., reflected in the empirically
observed increase of the variation coefficient $V=\theta/\bar{v}$ and
offers a safety-oriented interpretation of the capacity drop, i.e.,
the significant reduction of traffic flow when it becomes unstable
\cite{Hall1,Daganzo-ST,Kerner-sync}.

Variance-driven time headways can also qualitatively explain the
distribution of times-to-collision, which is surprisingly invariant
with respect to density changes (compared to distance, time gap, or
velocity distributions). Times-to-collision are, therefore, not only
an interesting measure for traffic safety, but also a meaningful
variable of behaviorally-oriented traffic models based on the physical
approach of invariants \cite{HDM,Lenz-Wagner}.

The variance-driven increase of the safety time gap $T$ may also be
seen as an alternative to a frustration-driven increase of $T$ after a
long time in congested traffic
\cite{IDMM,Tilch-TGF99,Neubert-TGF99,Leiche}. Moreover, it potentially
overcomes the criticism of traffic models with a fundamental diagram
by Kerner \cite{Kerner-book}, as it causes a pronounced platooning
effect when traffic flow is perturbed or unstable. This guarantees a
wide gap distribution which is the main prerequisite to reproduce the
wide scattering of flow-density data in congested traffic
\cite{Katsu03,Kerner-Rehb96-2}.

Previous explanations of the wide scattering of flow-density data
include stochastic effects \cite{Gipps81,Brackstone-hist}, and
sustained non-equilibrium states caused by dynamic instabilities such
as stop-and-go traffic
\cite{IDMM}.  Stochastic terms are included in most cellular-automaton
traffic models \cite{Nagel-S,Knospe-CA-test-04} and also in some
car-following models, e.g., in the Gipps-model \cite{Gipps81} or in
recent car-following models proposed by Kerner \cite{Kerner-mic} or
Wagner \cite{Wagner-fluct}. Another explanation of the scattering is
based on the heterogeneity of vehicles (such as cars and trucks) and
driving styles (such as defensive or aggressive)
\cite{GKT-scatter,Daganzo-ST,Helb-crit}. However, while all these
effects can possibly account for the observed variations of time headways
and times-to-collision, at least for a given traffic density, the
scattering of flow-density data would be smaller than observed due to
the averaging implied in aggregating single-vehicle data to, e.g.,
one-minute data \cite{Katsu03}.

In the next section, we will introduce the mechanism of
variance-driven time headways (VDT) in terms of a 'meta-model' which
can be applied to a wide range of car-following
models. Section~\ref{sec:fluct} introduces a general method to add
fluctuations to car-following models. In Sec.~\ref{sec:results}, we
apply the VDT mechanism to three microscopic traffic models and
compare 'virtual-detector data' directly with empirical findings. We
found that {\it this simple mechanism can semiquantitatively explain
all the microscopic and macroscopic empirical findings mentioned
above.}  In the concluding Sec.~\ref{sec:discussion}, we discuss the
effects of the VDT mechanism in terms of a spontaneous formation and
decay of vehicle platoons, and point to applications in the field of
traffic control and driver-assistance systems.

\section{\label{sec:model}Variance-driven adaptation of the time headway}
%
We will formulate the variance-driven time headways (VDT) model in
terms of a meta-model to be applied to any car-following model where
the time headway $T_0$ for equilibrium traffic can be expressed by a
model parameter or a combination of model parameters.

The basic assumption of the VDT is that smooth traffic flow allows for
lower values of the time headway than disturbed traffic flow where the
actual time headway
\begin{equation}
\label{T}
T=\alpha_T T_0
\end{equation}
is increased with respect to $T_0$ by a factor $\alpha_T\ge
1$. Furthermore, we characterize disturbed traffic flow (such as
stop-and-go traffic) by relatively high values of the velocity
differences between following vehicles. Since a driver in vehicle
$\alpha$ must assess the heterogeneity of traffic flow {\it in situ},
any measure for the heterogeneity may only depend on the immediate
environment. One of the simplest measure satisfying this requirement
is the {\it local variation coefficient}
\begin{equation}
\label{Varcoeff}
V_n=\frac{\sqrt{\theta_{n}}}{\bar{v}_{n}}
\end{equation}
where the local velocity average
\begin{equation}
\label{barv}
\bar{v}_{n}=\frac{1}{n} \sum \limits_{i=0}^{n-1} v_{\alpha-i},
\end{equation}
and the local variance
\begin{equation}
\label{sv}
\theta_{n}=\frac{1}{n-1} \sum \limits_{i=0}^{n-1} 
(v_{\alpha-i}-\bar{v}_{n})^2
\end{equation}
are calculated from the own velocity $v_{\alpha}$ and the velocities
of the $(n-1)$ predecessors $(\alpha-i), i=1, \ldots, n-1$.
For the sake of simplicity we will skip the vehicle index $\alpha$
here, and in
all subsequent equations. In this work, we will set $n=5$ in most
cases, i.e., the adaptation of the drivers is assumed to depend on the
own velocity and the velocities of the four nearest vehicles in front.

This quantity can be empirically determined if single-vehicle data are
available. Figure \ref{figEmp}(c) shows an example for the Dutch
freeway A9 between Haarlem and Amsterdam. Notice that, for a given
local density $\rho$, the variation coefficient $V_n=\sqrt{A}$ is
related to the variance prefactor $A(\rho)$ introduced in the
macroscopic gas-kinetic-based traffic (GKT) model \cite{GKT}.

Let us now assume that the multiplication factor $\alpha_T$ of the
time headway is adapted instantaneously to a traffic situation
according to
\begin{equation}
\label{alphaT}
\alpha_T
=\min\left(1+\gamma V_n, \alpha_T\sup{max} \right).
\end{equation}
Here, $\alpha_T\sup{max}$ denotes the maximum multiplication factor
for the time headway for traffic flows of maximum unsteadiness, and
\begin{equation}
\label{gamma}
\gamma=\frac{1}{T_0}\abl{T}{V_n}
\end{equation}
the sensitivity of the time headway to increasing velocity variations.

In summary, Eqs. \refkl{Varcoeff} - \refkl{alphaT} imply that the
necessary time headway for safe driving depends on the velocity
variance of the surrounding traffic. This proposition of
variance-driven time headways (VDT) can be applied to any
time-continuous car-following model in which the time headway can be
expressed by a model parameter or a combination of parameters. Some
examples are the optimal-velocity model (OVM) \cite{Bando}, the
velocity-difference model (VDIFF) \cite{Jiang-vDiff01}, the
intelligent-driver model (IDM) \cite{Opus}, or the Gibbs model
\cite{Gipps81}.
 
The VDT has three parameters, namely the number $n$ of vehicles used
to determine the local velocity variance, the maximum multiplication
factor $\alpha_T$ by which the time headway is increased compared to
perfectly smooth traffic, and the sensitivity $\gamma$.  For the
special case $n=2$, the VDT acceleration depends only on the velocity
difference to the immediate predecessor, i.e., one obtains a simple
car-following model depending only on the immediate predecessor (at
least, if this is the case for the underlying car-following
model). However, the model yields more realistic results for a larger
number of vehicles, therefore we will assume $n=5$ in all simulations
(see Table \ref{tab:VDT}). Larger values for $n$ will not change the
dynamics significantly.

The parameters $\alpha_T$ and $\gamma$ can be determined from
empirical data of the time-headway distribution for free and congested
traffic, and from the observed maximum variation coefficient
$V_n\sup{max}$.  Figure \ref{figEmp} shows these data for the Dutch
freeway A9 from Haarlem to Amsterdam
\cite{Tilch-TGF99}. Figure \ref{figEmp}(a) shows that the
locations of the maxima 
(modal values) of the time-headway distributions for free and
congested traffic differ by a factor of about two. We therefore 
set $\alpha_T\sup{max}=2$ in all simulations.
The parameter $\gamma$ can be determined by the approximate relation
\begin{equation}
\label{gamma_emp}
\gamma \approx \frac{\alpha_T\sup{max}-1}{V_n\sup{max}}.
\end{equation}
From Fig. \ref{figEmp}(c) we see that $V_n\sup{max}$ is slightly below
0.2, so we set $\gamma=5$ in all simulations.

\begin{table}
\begin{tabular}{lr}
Parameter        & Value \\[1mm] \hline
\\[-2mm]
Number $n$ of vehicles for determining $\theta$ \quad & 5\\
Time-headway multiplication factor & \\ 
in unsteady traffic $\alpha_T\sup{max}$ & 2.2 \\
Sensitivity $\gamma$ & 4.0 \\[2mm]
\end{tabular}

  \caption{\label{tab:VDT} Model parameters of the VDT approach used
  throughout this paper for all simulated car-following models.  The
  strength of the acceleration noise
  (cf. Sec.~\protect\ref{sec:fluct}) was set to
  $0.1\,\text{m}^2/\text{s}^3$.}
\end{table}

\section{\label{sec:fluct}Acceleration noise}
%
Fluctuating forces in microscopic traffic models are used to globally
describe all influences that are not modeled explicitly such as
imperfect estimation capabilities \cite{HDM}, lack of attention, or
simply the fact that drivers do not always react identically to a
given traffic situation.  Fluctuation terms are part of nearly all
cellular automata (the most popular example being the
Nagel-Schreckenberg model \cite{Nagel-S} and models derived from it
\cite{Chowdhury-CA-Opus}), but are less commonly used in
time-continuous car-following models.

Since the VDT is essentially based on fluctuations of the velocity, it is
to be expected that purely deterministic underlying models yield
unrealistic results due to the lack of an initial source triggering
the fluctuations. Therefore, we consider additional acceleration
fluctuations when applying the VDT to a deterministic
model. 

For simplicity, we will just add a white (independent and
$\delta$-correlated) noise term
\cite{FPE-03-preprint}
to the deterministic car-following acceleration
$a\sup{(det)}_{\alpha}$ according to
\begin{equation}
\label{stoch}
\dot{v}_{\alpha} = a\sup{(det)}_{\alpha}(t) + \sqrt{Q} \, \xi_{\alpha}(t).
\end{equation}
Here, $Q$ denotes the fluctuation strength (cf. Table
\ref{tab:VDT}), and
the white noise $\xi(t)$ is assumed to be unbiased and
$\delta$-correlated:
\begin{equation}
\label{Qa}
\erw{\xi_{\alpha}}=0, \ \ \erw{\xi_{\alpha}(t)\xi_{\beta}(t')}
= Q \delta_{\alpha\beta}\delta(t-t').
\end{equation}
The Kronecker symbol $\delta_{\alpha\beta}$ is 1, if $\alpha=\beta$
and zero otherwise, while the Dirac function $\delta(t)$ is defined by
$\int_{-\infty}^{\infty} \delta(t') \, dt'=1$ and $\delta(t)=0$ for $t
\neq 0$. To clarify the effects of the fluctuation term on the velocity, we
note that
\begin{itemize}
\item[(i)] in the
absence of a deterministic acceleration, Eq. \refkl{stoch} leads to
velocities $v_{\alpha}(t)$ fluctuating stochastically around the
initial velocity $v_{\alpha}(t_0)$ with a linearly-in-time increasing
variance (random walk),
\begin{equation}
\label{randomWalk}
\theta_{\alpha}=Q(t-t_0),
\end{equation}
\item[(ii)] under the linearized deterministic (relaxational) dynamics 
$ a\sup{det}_{\alpha}=(v_0-v)/\tau $ (where $v_0$ is the desired
velocity, $v$ the actual velocity, and $\tau$ the acceleration
relaxation time), the velocity variance of the stationary state is
given by the fluctuation-dissipation result \cite{Gardiner}
\begin{equation}
\label{fluctdiss} 
\theta=Q\tau.
\end{equation}
\end{itemize}
In the explicit numerical velocity update from time $t$ to $t+\delta
t$,
\begin{equation}
\label{stochUpdate}
v_{\alpha}(t+\delta t)=v_{\alpha}(t)
+a_{\alpha}\sup{det} \delta t + \eta_t \sqrt{Q\delta t},
\end{equation}
we have implemented the stochastic term by the additive contribution
$\eta_t \sqrt{Q\delta t}$, where the $\{\eta_t\}$ are statistically
independent realizations of Gaussian distributed random numbers with
zero mean and unit variance \cite{Gardiner}.

The velocity update according to \refkl{stochUpdate} is numerically
consistent in the stochastic sense.  More precisely, we have
considered the 'numerical' velocity distribution function
$F\sup{num}(v,t)=\mbox{Prob}(v_{\alpha}(t) \le v)$ obtained from many
simulations with different seeds for the pseudo-random number
generator at a given time $t$. We have then compared the numerical
distribution function with the theoretical distribution
$F(v,t)=\int_{-\infty}^{v} dv' P(v', t)$ where $P(v,t)$ is the
solution to the Fokker-Planck equation corresponding to
Eq. \refkl{stoch},
\begin{equation}
\label{FPE}
\ablpart{P(v_{\alpha},t)}{t} 
+ a_{\alpha}\sup{(det)} \ablpart{P}{v_{\alpha}} = \frac{Q}{2}
\ablparttwo{P}{v_{\alpha}}.
\end{equation}
It turned out that the deviations between $F\sup{num}(v,t)$ and
$F(v,t)$ are of the order ${\cal O}(\delta t)$. Notice that this means
that, for sufficiently small time steps $\delta t$, the result is
independent from $\delta t$ and agrees with the analytic solution to
the stochastic differential equation \refkl{stoch}.

We have checked this for the random walk and linear relaxation
scenarios mentioned above and found a very good agreement between the
numerical results from the update scheme \refkl{stochUpdate} and the
analytical results \refkl{randomWalk} and \refkl{fluctdiss},
respectively.  (see Ref. \cite{FPE-03-preprint} for a more detailed
discussion).

In summary, Eqs. \refkl{stoch} - \refkl{stochUpdate} can be considered
as a general approach to add acceleration noise to time-continuous
deterministic car-following models. Clearly, $\delta$-correlated noise
terms are unrealistic in many respects. Therefore, we have also
simulated more realistic time-correlated and multiplicative noise,
which clearly describes the human origins of acceleration noise better
\cite{HDM}, but we found no qualitative difference.

\begin{figure}
  \begin{center}
    \includegraphics[width=60mm]{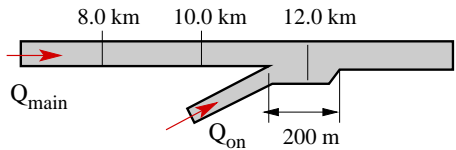}
  \end{center}

  \caption{\label{figInfra} Simulated infrastructure and positions of
  virtual detectors. The units of locations are measured in
  kilometers.  }
\end{figure}

\section{\label{sec:results}Simulation results}
%

In the following, we will apply the VDT to three car-following models,
namely the intelligent-driver model (IDM) \cite{Opus}, the
optimal-velocity model (OVM) \cite{Bando}, and the velocity-difference
model \cite{Jiang-vDiff01}, which augments the OVM by a term
proportional to the velocity difference. We will also simulate
heterogeneous traffic consisting of a mixture of these models.

For the purpose of reference and in order to discuss the coupling to
the VDT, we shortly present the model equations, i.e., the
acceleration functions, of these models.

The IDM acceleration $\dot{v}\sub{IDM}(s,v,\Delta v)$ of a vehicle as
a function of the (net) distance $s$ to the predecessor, the velocity
$v$, and the velocity difference $\Delta v$ (positive when
approaching) is given by
\begin{equation}
\label{IDM}
\dot{v}\sub{IDM} 
 = a \left[ 1-\left(\frac{v}{v_0}\right)^4 
              - \left(\frac{s^*}{s}\right)^2 \right]
\end{equation}
with the 'desired dynamical distance'
\begin{equation}
s^*=s_0+vT+\frac{v \Delta v}{2\sqrt{ab}}.
\end{equation}
The acceleration of the optimal-velocity and velocity-difference
models is given by
\begin{equation}
\label{OVM}
\dot{v}\sub{OVM}  = \frac{v\sub{opt}-v}{\tau} - \lambda \Delta v
\end{equation}
where $\lambda=0$ for the OVM and the 'optimal velocity' is given by
\begin{equation}
v\sub{opt}=\frac{v_0}{2}\left[ \tanh\left(\frac{s}{L}
    -\beta\right) - \tanh(-\beta) \right].
\end{equation}
The coupling \refkl{T} of the VDT to the IDM is simple, since the
desired time headway $T$ is already an IDM parameter. We used
$T=T_0=0.7$ s as minimum value which can be increased up to $T=1.54$ s
corresponding to $\alpha_T\sup{max}=2.2$ (cf. Table \ref{tab:VDT}).
To find an appropriate coupling of the VDT to the OVM and VDIFF
models, we note that the parameter $L$ defines a typical interaction
range and, consequently, the desired time headway $s/v\sub{opt}(s)$ is
essentially proportional to $L/v_0$ in these models.  Therefore, we
coupled the VDT to the OVM and the VDIFF models by setting $L=L_0
\alpha_T$ with $\alpha_T$ according to Eq. \refkl{alphaT}.

In order to distinguish between freely moving and following vehicles,
we need at least two vehicle types ('cars' and 'trucks') with
different desired velocities $v_0$. For all models, we have set
$v_0=35$ m/s for 'cars', and $v_0=25$ m/s for 'trucks' and simulated a
truck percentage of 20\%.  Notice that $v_0$ is a common parameter of
all three models.  Because we want to introduce as little complexity
as possible, we did not distinguish cars and trucks with respect to
other parameters.  Furthermore, we used the same vehicle length
$l\sub{veh}=5$ m for both vehicle types in all simulations.

The remaining IDM parameters are the minimum gap $s_0=3$ m, the
acceleration $a=1\, \msii$, and the comfortable deceleration
$b=1.5\,\msii$. For the OVM, we used the relaxation time $\tau=0.4$ s
and the form factor $\beta=1$. Furthermore, we set the interaction
length $L_0=13$ m for cars and $L_0=10$ m for trucks. Thus, the
effective time headway is about the same for both types.  For the
VDIFF, we used the same values for $L_0$ and $\beta$ as for the
OVM. Furthermore, we used $\tau=2$ s and the sensitivity coefficient
$\lambda=1 \, {\rm s}^{-1}$.  For all models, we set the fluctuation
strength $Q=0.1 \, {\rm m}^2/{\rm s}^3$. For comparison, we simulated
also the deterministic VDT-IDM for which the fluctuation strength is
$Q=0$. The parameters were chosen such that the traffic dynamics was
comparable to the Dutch freeway A9 freeway data with respect to the
form of the fundamental diagram, capacity, and stability.

\begin{figure}
  \begin{center}
    \includegraphics[width=88mm]{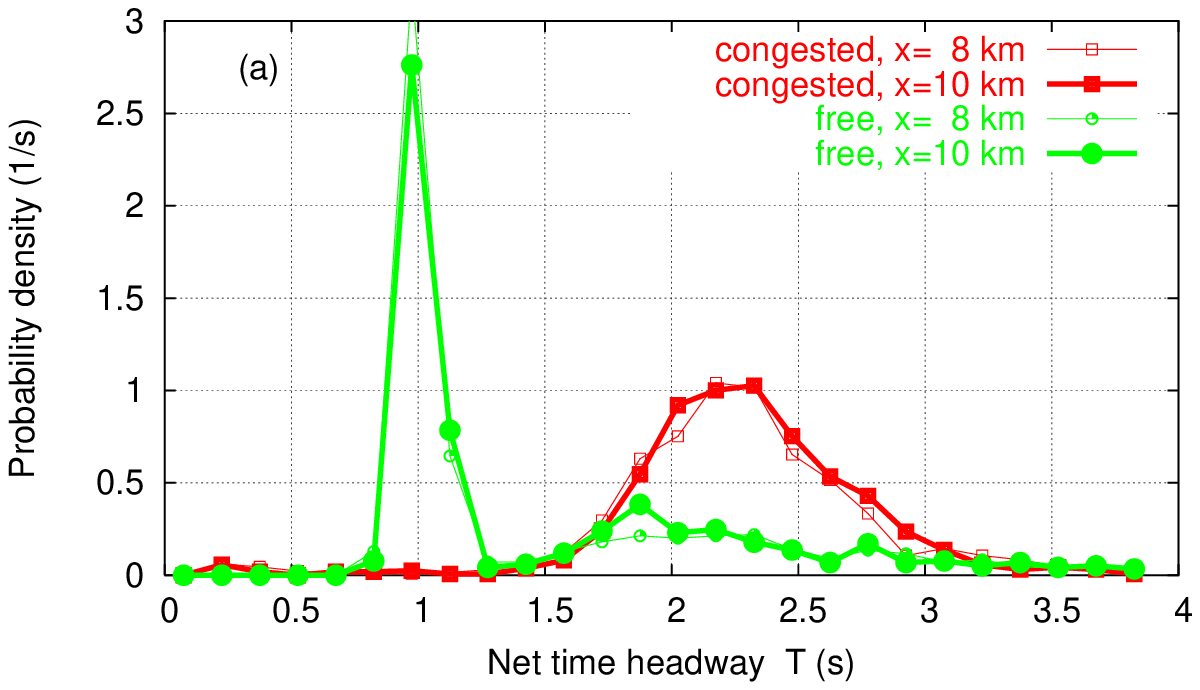}
    \includegraphics[width=88mm]{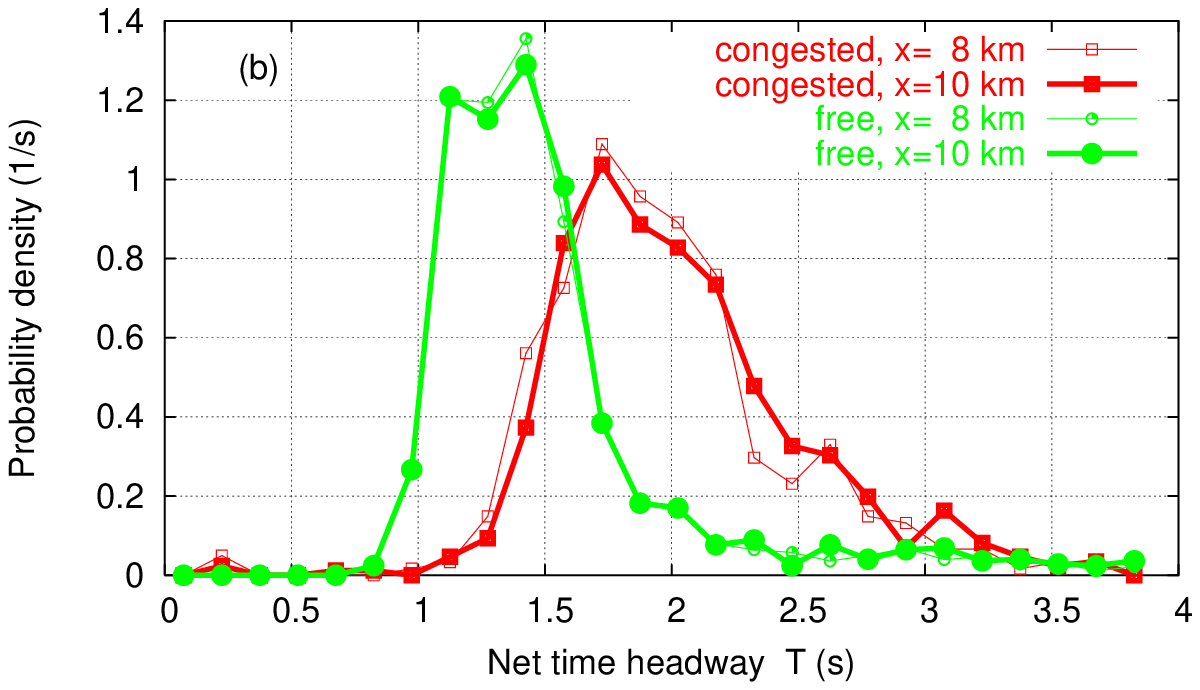}
    \includegraphics[width=88mm]{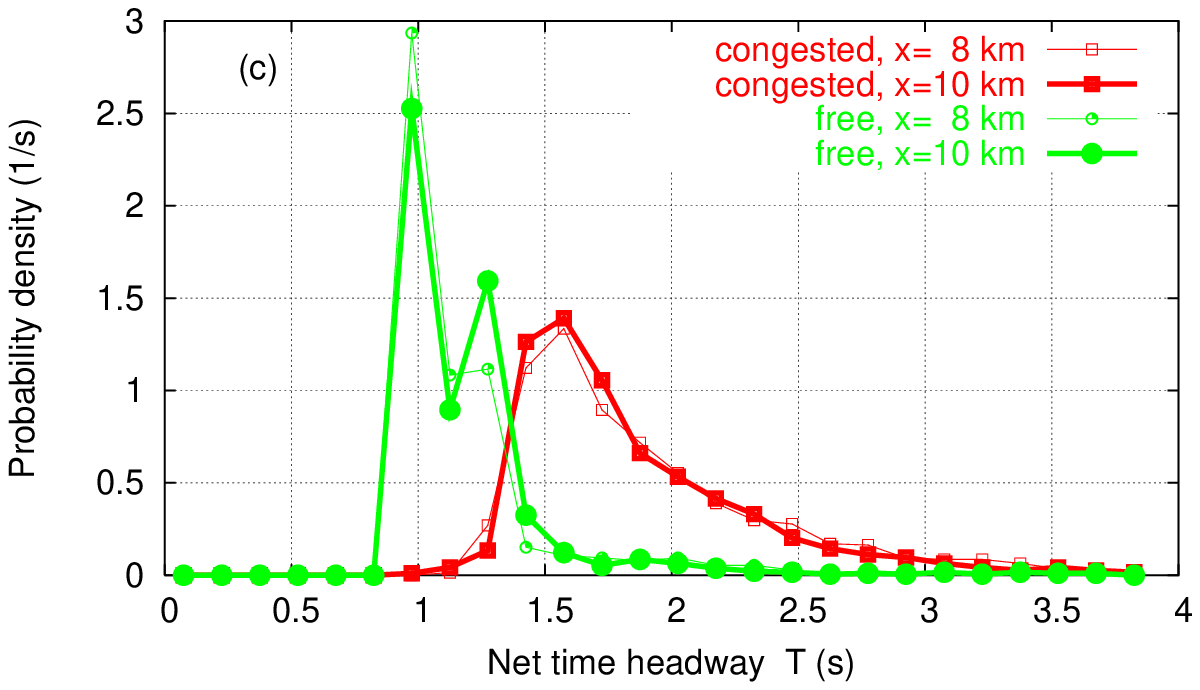}
  \end{center}

 \caption{\label{figT} Distribution of the net time headways of cars
 following any kind of vehicle (cars or trucks) obtained from
 single-vehicle data of 'virtual detectors' at various positions for
 simulations of the VDT with (a) the IDM; (b) the OVM; (c) the VDIFF
 model.}
\end{figure}

\begin{figure}
  \begin{center}
    \includegraphics[width=88mm]{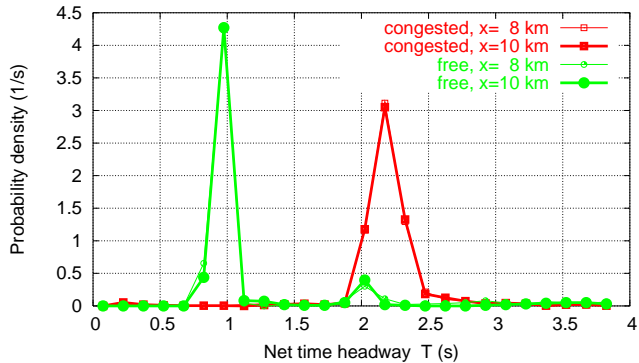}
  \end{center}

 \caption{\label{figTdet} Distribution of the net time headways of
 cars for the deterministic VDT-IDM.  }
\end{figure}

We have simulated a single-lane road section of total length 15 km
with an on-ramp of length $L\sub{rmp}=200$ m located at
$x\sub{rmp}=12$ km (Fig. \ref{figInfra}) from which a constant flow of
400 vehicles/h merges to the main road. To keep matters simple, we
have avoided explicit modeling of the merging of ramp vehicles to the
main road. Instead, we have inserted the ramp vehicles centrally into
the largest gap within the 200 m long ramp section.  In order to
generate a sufficient velocity perturbation in the merge area, the
speed of the accelerating on-ramp vehicles at the time of insertion
was assumed to be 50\% of the velocity of the respective front
vehicle. It turned out that the perturbations induced by the slower
merging vehicles were crucial: When simulating merges with the same
velocity as the main road vehicles, the onset of traffic breakdown was
markedly delayed indicating the role of perturbations for traffic
optimization (see Sec. \ref{sec:discussion} below).

We initialized the simulations with very light traffic of density
$\rho=3$ vehicles/km and an initial velocity of 100 km/h. The details
of the initial conditions, however, are not relevant unless they lead
to an immediate breakdown of traffic flow.  To generate congestion, we
have increased the inflow of vehicles to the main lane linearly from
300 vehicles/h at $t=0$ s to 3000 vehicles/h at $t=2400$
s. Afterwards, we decreased the inflow linearly to 300 vehicles/h
until $t=4800$ s.  In case the inflow exceeded capacity, we delayed
the insertion of new vehicles at the upstream boundary.

The update time step of the numerical integration scheme was $\delta
t=0.05\:\mathrm{s}$ for all models. Runs with smaller time steps
yielded essentially the same results.

\subsection{\label{sec:res-mic}Time-headway distribution}
%
Empirical investigations of single-vehicle data have shown that the
distributions of net time headways differ markedly in free and
congested traffic situations \cite{Tilch-TGF99,Kno02-data}, see
Fig. \ref{figEmp}.

To enable direct comparisons with experimental work, we have
implemented 'virtual detectors' at $x=8$~km and 10~km
(cf. Fig.~\ref{figInfra}) recording the passage time $t_{\alpha}$, type
(car or truck) and velocity $v_{\alpha}$ of each vehicle $\alpha$
crossing the detector.  We estimated the net time headway $T_{\alpha}$
by the time interval between the passage of the rear bumper of the
preceding vehicle $(\alpha-1)$ and the front bumper of the vehicle
under consideration,
\begin{equation}
\label{Talpha}
T_{\alpha}=t_{\alpha}-t_{\alpha-1}-\frac{l_{\alpha-1}}{v_{\alpha-1}}.
\end{equation}

Figure~\ref{figT} shows the simulated distribution of $T_{\alpha}$ for
the faster vehicle type ('cars' following any vehicle type) separately
for free traffic ($v_{\alpha}>15$ m/s) and congested traffic
($v_{\alpha} \le 12$ m/s). We have obtained the following main
results:
\begin{itemize}
\item The modal value
(location of the maximum of the probability density) of the time
headway is markedly higher (about twice as high for the VDT-IDM) in
congested traffic compared to free traffic.
\item The values for $T_{\alpha}$ form a broad and
asymmetrical distribution.
\item The different
underlying models, particularly deterministic and stochastic variants,
yielded the same qualitative results. Remarkably, the velocity
variance depends only weakly on the noise.
\end{itemize}
Notice that the VDT only prescribes that the time headway increases
with the variance. The dependence of the variance (and thus the
average time headway) on the traffic situation results from the
traffic dynamics. Moreover, all statistical data were obtained from
identical vehicles (the 'cars'). Since all cars have the same unique
equilibrium relation between velocity and net distance $s$, the wide
and bimodal distributions are an interesting result, particularly for
the deterministic case.

\begin{figure}
  \begin{center}
    \includegraphics[width=88mm]{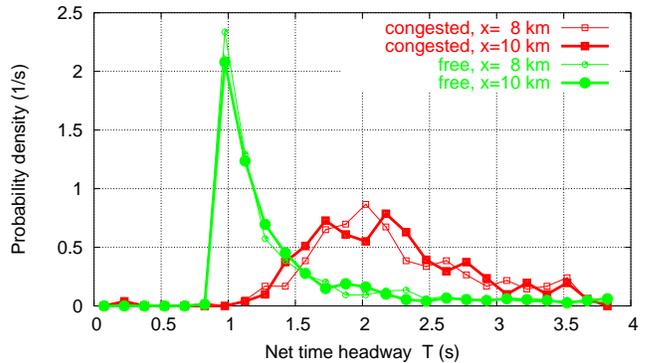}
  \end{center}

 \caption{\label{figHet} Distribution of time headways for the VDT
 with a mix of 1/3 IDM, 1/3 OVM, and 1/3 VDIFF vehicles (truck
 percentage 20\% in each model).  }
\end{figure}

Nevertheless, the peaks of the simulated distributions are higher and
sharper that those in the empirical data (cf. Fig.~\ref{figEmp}) The
lacking quantitative agreement in this case can be explained by the
wide variation of individually preferred time headways of drivers,
i.e., in the variations of the driving style \cite{Katsu03}, which
was neglected in our simulation for reasons of simplicity.
To test this assumption, we have simulated a mix of all three models.
The resulting time-headway distribution for congested traffic and the
variance as a function of the density reproduces the observed data
nearly quantitatively as shown in Fig.~\ref{figHet}.

For the sake of simplicity, we will, however, not incorporate
heterogeneity in the rest of this work.

\subsection{Times-to-collision}
%
Further dynamic microscopic aspects of traffic can be captured by the
'times-to-collision' (TTC) $\tau_{\alpha}\sup{TTC}$ which is defined as
the time interval after which a vehicle $\alpha$ would collide with
its predecessor provided no deceleration would take place. Negative
TTC values denote 'virtual' collisions in the past. We extract the
value of $\tau\sup{TTC}_{\alpha}$ from the single-vehicle data by the
relation
\begin{equation}
\label{TTC}
\tau\sup{TTC}_{\alpha} = \frac{s_{\alpha}}{v_{\alpha}-v_{\alpha-1}}
 \approx \frac{T_{\alpha} v_{\alpha-1}}{v_{\alpha}-v_{\alpha-1}},
\end{equation}
where $s_{\alpha}$ is the net distance inferred from
the data.

\begin{figure}
  \begin{center}
    \includegraphics[width=88mm]{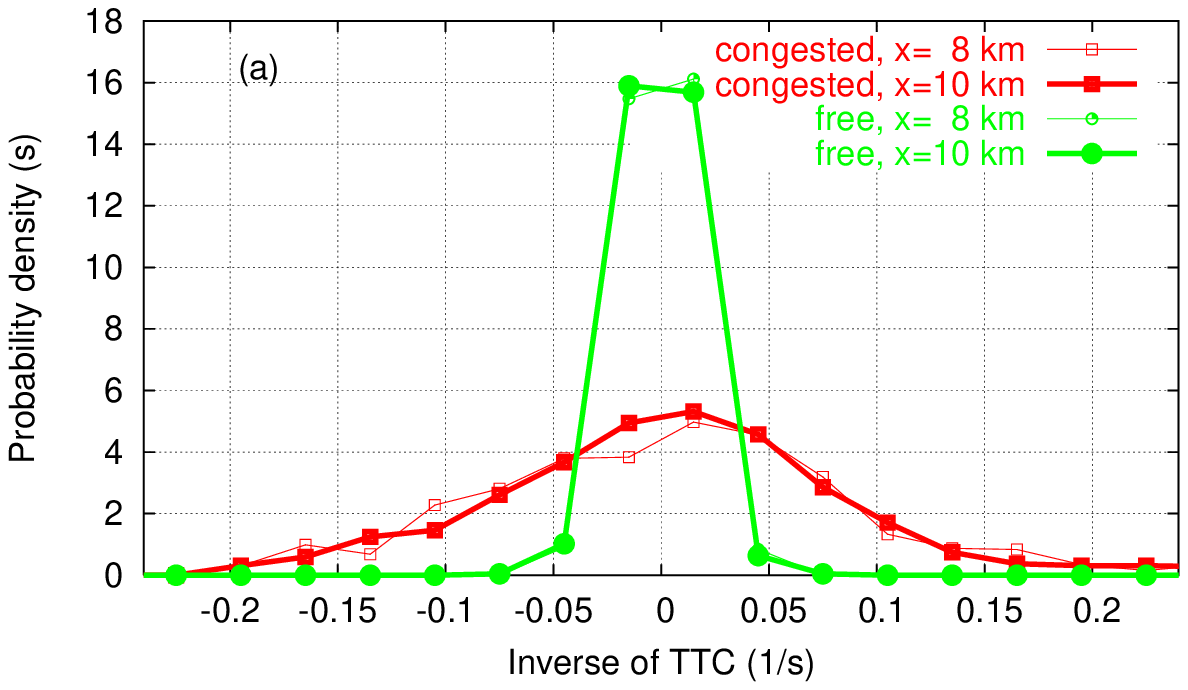}
    \includegraphics[width=88mm]{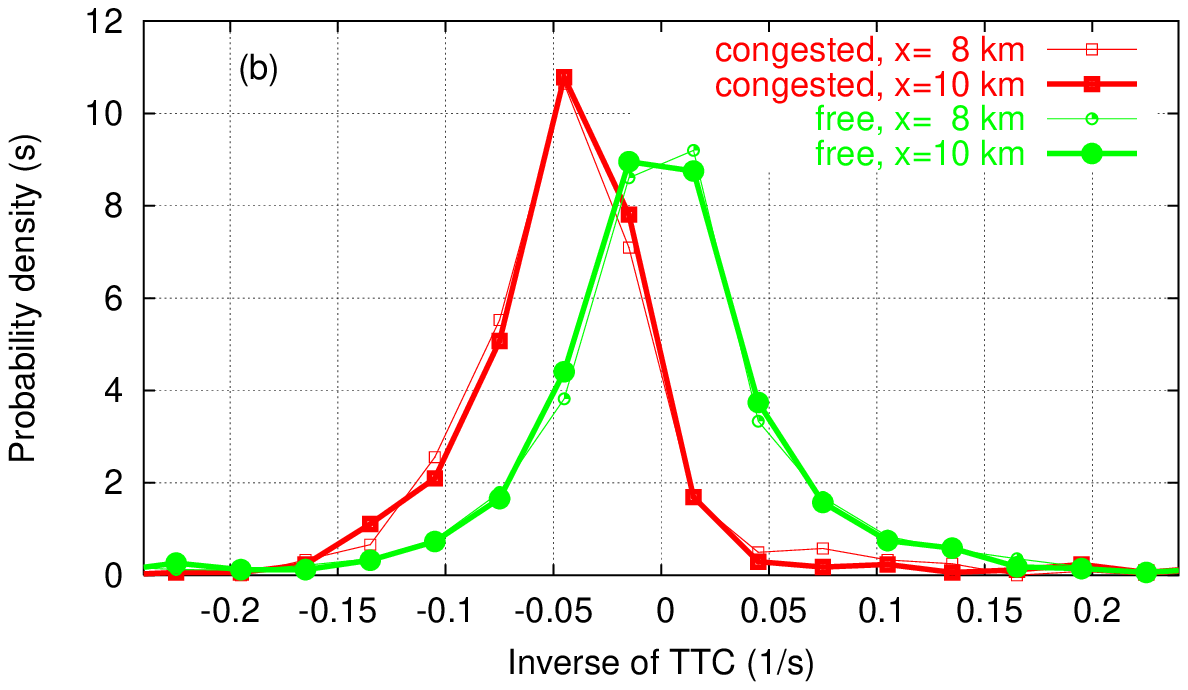}
    \includegraphics[width=88mm]{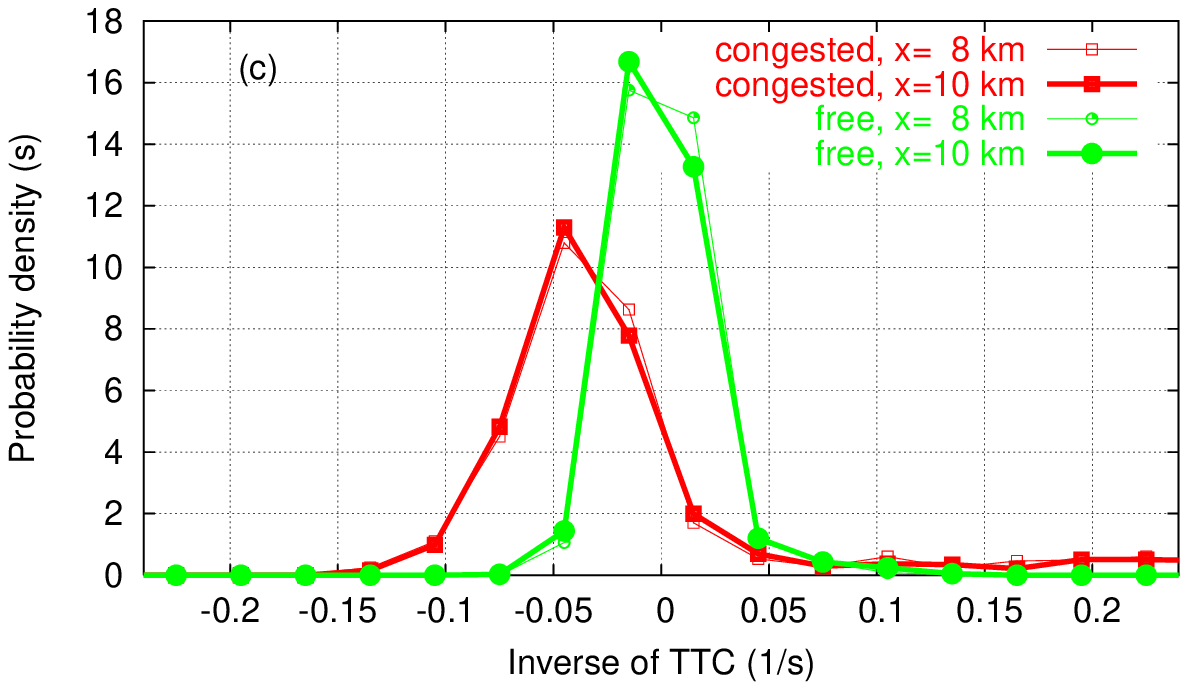}
  \end{center}

 \caption{\label{figTTC1} Distribution of the inverse
 $(v_{\alpha}-v_{\alpha-1})/s_{\alpha}$ of the times-to-collision as
 calculated from single-vehicle data of virtual detectors for the VDT
 using (a) the IDM; (b) the OVM; (c) the VDIFF.  }

\end{figure}

Instead of plotting distributions of the TTC directly, we present, in
Fig. \ref{figTTC1}, distributions of the inverse of the TTC, which we
denote as 'relative approaching rate'
\begin{equation}
\label{invTTC}
r_{\alpha}=1/\tau\sup{TTC}_{\alpha}.
\end{equation} 
This has the advantage that the interesting range of small positive
and negative values of the TTC is magnified and there is no divergence
for the equilibrium state, i.e., $v_{\alpha}=v_{\alpha-1}$. We have obtained
the following main results:
\begin{itemize}
\item Both in the models and in the empirical data
(cf. Fig. \ref{figEmp}), the distribution is nearly symmetrical with
respect to positive and negative values of $r_{\alpha}$.
\item For the OVM and VDIFF, the width of the distribution
is nearly the same for free and congested traffic, in agreement with
the empirical data, while the variance of the IDM distribution is too
small for free traffic.
\item For the IDM, the peaks of the distribution are located 
at $r_{\alpha}=0$ (i.e., equilibrium traffic is the most probable
state, in agreement with the data) while the peak for congested
traffic is shifted towards negative times-to-collision in the other
two models.
\end{itemize}
The main difference of the IDM with respect to the other models are
the order of magnitude of the accelerations. While the IDM
acceleration (whose order is given by the parameters $a$ and $b$) did
not exceed $\pm 4 \msii$, the OVM acceleration varied between $- 16
\msii$ and $7\,\msii$, and the VDIFF accelerations between $-
9\,\msii$ and $5\,\msii$. Since the determination of $r_{\alpha}$ from
single-vehicle data via the Eqs. \refkl{TTC} and \refkl{invTTC} is
only exact for zero accelerations, this is a possible explanation for the
asymmetry in the simulation. 

\section{\label{sec:discussion}Discussion}
%
In the variance-driven time headway (VDT) model put forward in this
paper, the desired safety time headway is a dynamic variable
increasing with the local velocity variance. This provides a mechanism
for a spontaneous formation and decay of long-lived but non-permanent
platoons: If traffic flow is stable, initial velocity differences
decrease, leading to decreased values of the local variance and
thereby to low values of the desired time headway and a high dynamic
road capacity. Because of the conservation of the vehicle number, this
automatically leads to platoons and to large gaps in front of the
slowest vehicles ('trucks').  For sufficiently high traffic demands
the short time gaps may result in unstable traffic flow, leading to
higher values of the variance. This, in turn, causes spontaneous
braking maneuvers of the drivers which further increase the velocity
variance. Finally, this breaks up the whole platoon resulting in a
traffic breakdown with a distinct capacity drop.

\begin{figure}
  \begin{center}
    \includegraphics[width=70mm]{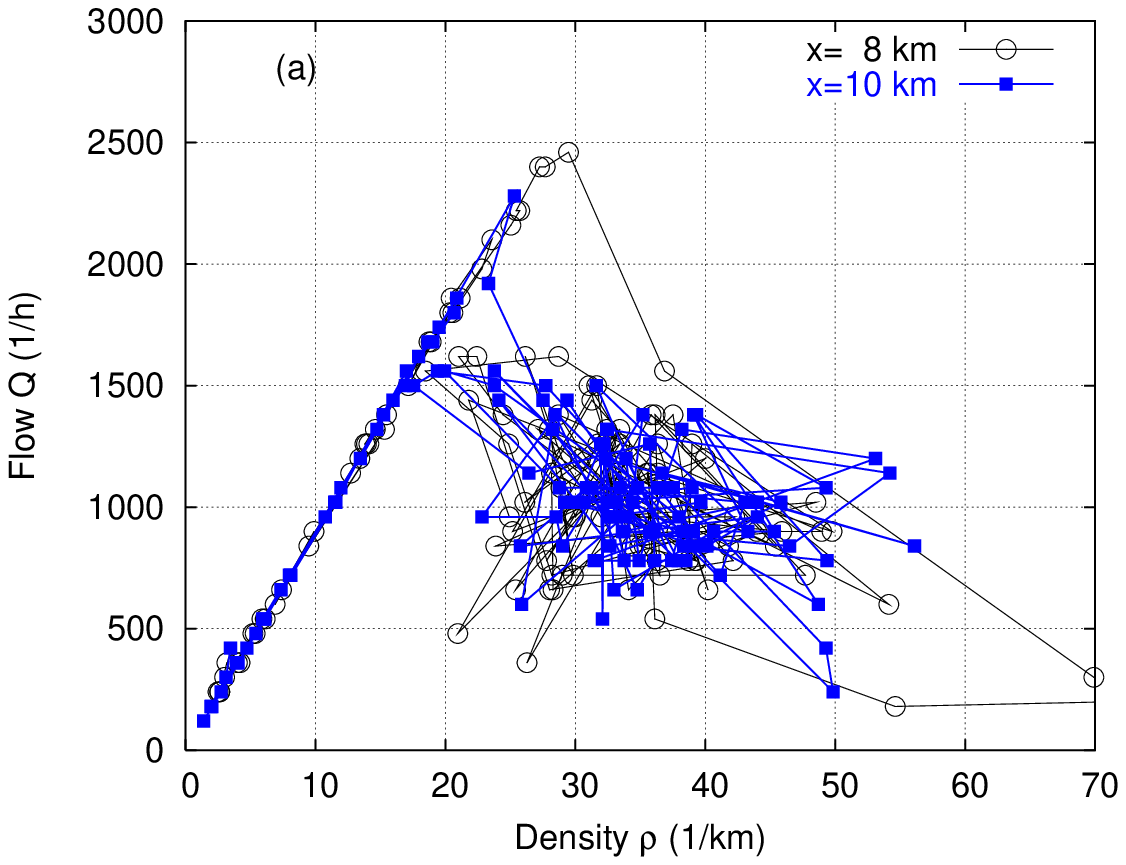}
    \includegraphics[width=70mm]{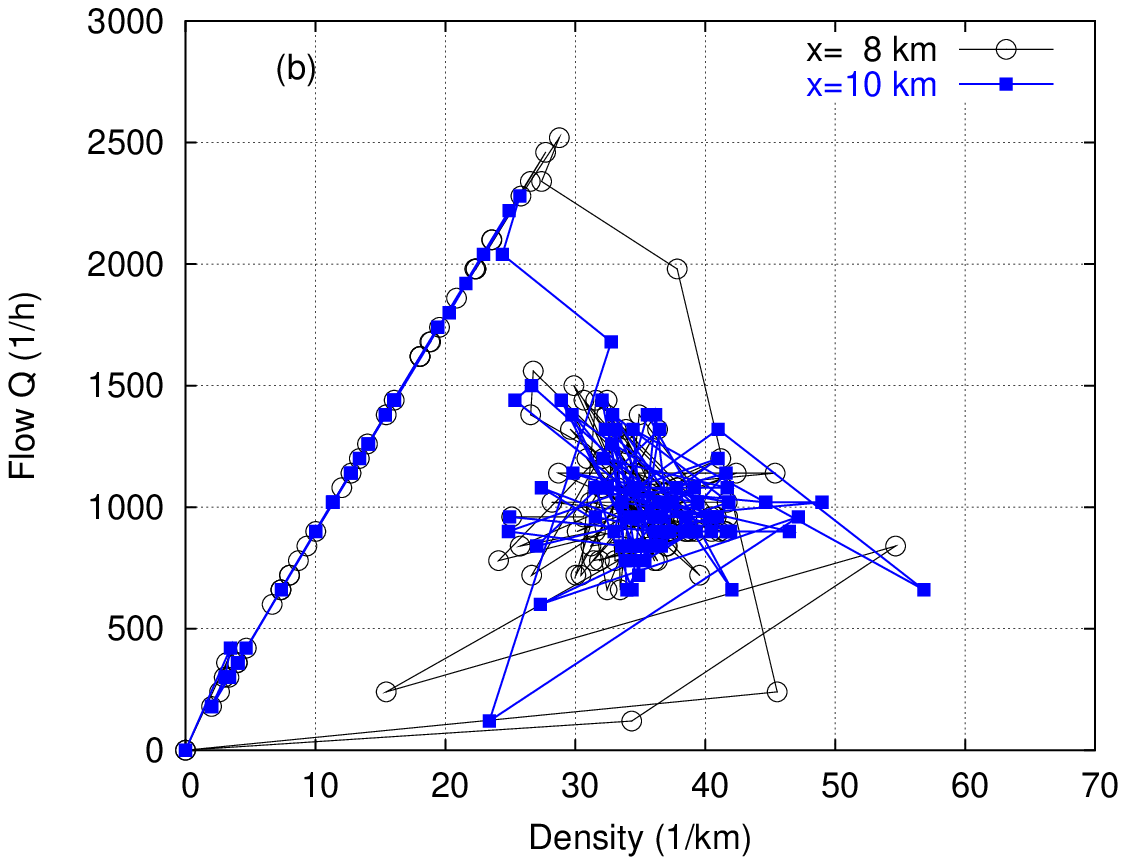}
    \includegraphics[width=70mm]{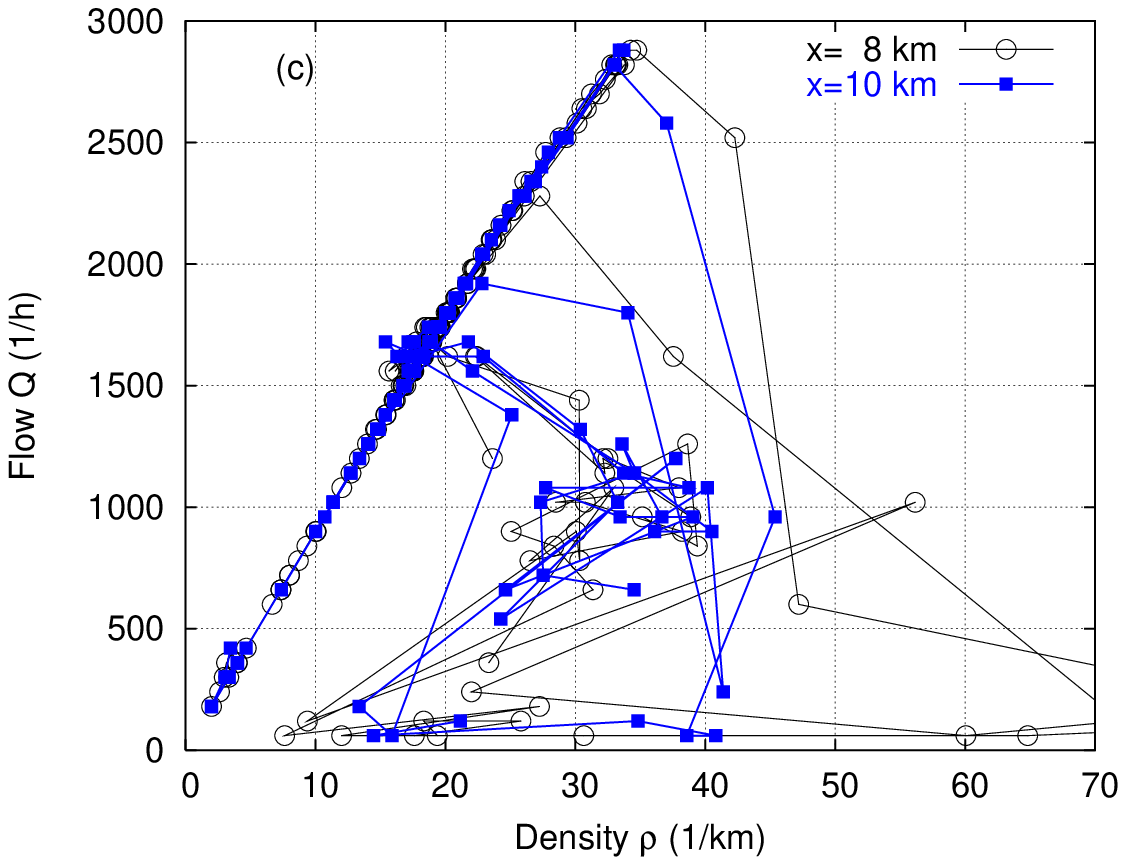}
  \end{center}

 \caption{\label{figFund} Simulated flow-density data of the VDT
 applied to the IDM at two 'virtual detectors' with a sampling
 interval of $T_\mathrm{aggr}=60$~s (a) with fluctuations; (b) without
 fluctuations; (c) without fluctuations and on-ramp vehicles merging
 with the speed of the vehicles on the main road rather than half of
 it as in (a) and (b).}
\end{figure}

Our intention is to propose a simple model for this variance-driver
mechanism.  Therefore, we have neglected, e.g., finite reaction times
or more elaborated concepts of anticipation which are contained, for
example, in the human driver model (HDM) \cite{HDM}.  Furthermore, we
have modeled fluctuations in the simplest possible way.

In the following, we want to discuss our main results in the light of
the 'VDT mechanism':
\begin{itemize}
\item The distribution of time headways in free traffic is broad and
asymmetric both in the deterministic and stochastic cases, although
all vehicles (cars and trucks) have the same time-headway
parameter. The reason is that the time headway depends dynamically on
the velocity variance. Consequently, even the deterministic
driver-vehicle units do not have a unique fundamental diagram.
\item The averaged time headway in congested
traffic is almost twice of that in free traffic (Figs. \ref{figEmp}
and \ref{figT}), which is related to the higher values of the velocity
variation coefficient for congested traffic compared to free traffic.
\item 
Apart from the IDM, the distribution of the relative approaching rates
(inverse of the times-to-collision) is nearly independent of whether
traffic is free or congested. This is a result of several antagonistic
effects: In congested traffic, the velocity correlation
$r_{v_{\alpha},v_{\alpha-1}}$ between neighboring vehicles, the
variation coefficient $V$, Eq.~\refkl{Varcoeff}, and the time headway $T$,
Eq.~\refkl{T}, are all higher than in free traffic. In the approximate
expression for the standard deviation of the relative approaching rate
defined in Eq.~\refkl{invTTC},
\begin{equation}
\sqrt{\erw{r_{\alpha}^2}} =
\frac{V(1-r_{v_{\alpha},v_{\alpha-1}})}{T},
\end{equation}
these three influences essentially cancel out each other for suitable
parameter choices.

\item On a macroscopic level, the VDT reproduces
 the wide scattering of data points in the flow-density diagram
 calculated from one-minute data (Fig.~\ref{figFund}), and the
 capacity drop at the transition from free to congested traffic
 (Fig.~\ref{figFund}(a)).
\end{itemize}

In presenting our simulation results, we have emulated the available
data analogously to real traffic data. For example, we did not use the
full information of all vehicle trajectories for plotting the
spatiotemporal dynamics of the velocity. Instead, we have restricted
ourselves to 'virtual detectors', 
as this approach allows a \textit{direct} comparison with
empirical traffic data.

We note that an understanding of the effects of the velocity variance
is crucial for devising measures to avoid traffic breakdowns: The VDT
feedback mechanism is triggered most likely near sources of sustained
velocity variations, for example in the merging, diverging, or weaving
zones near freeway intersections, but also the noise term plays a
role.  To illustrate that, we have introduced a sustained velocity
perturbation in all simulations of Sec. \ref{sec:results} and in
Figs.~\ref{figFund}(a) and (b) by letting accelerating ramp vehicles
merge with only half the velocity of the vehicles on the main road.
In simulations where the ramp vehicles entered with the speed of the
vehicles on the main road, we observed a markedly delayed traffic
breakdown occurring only after a traffic-flow peak near 3000
vehicles/h instead of 2500 vehicles/h, cf. Fig.~\ref{figFund}(c). The
subsequent breakdown, however, was more severe showing not only
'synchronized traffic' but also jammed traffic with nearly vanishing
flows.  Eliminating the noise term alone had a smaller effect
(Fig.~\ref{figFund}(b)).

Thus, it is essential to avoid merging and diverging maneuvers at high
velocity differences, e.g., by increasing the length of the
acceleration lane at on-ramps and off-ramps.  Another measure to reduce
the velocity variance are speed limits which can be simulated with the
VDT as well. Furthermore, since lane changes constitute another source
of velocity variance, we expect a strong coupling of lane changes to
the longitudinal dynamics by the VDT mechanism. In particular, a
multi-lane generalization of the VDT \cite{MOBIL-rostock} might yield
a fully quantitative explanation of bottleneck effects introduced by weaving
zones and off-ramps.

Finally, the distinct increase of the time headways after traffic
breakdown opens up vehicle-based options to increase traffic
performance and stability by means of adaptive-cruise control (ACC)
systems. Such driver-assistance systems, which accelerate and brake
automatically depending on the distance to the preceding vehicle and
its velocity, are already commercially available for some upper-class
vehicles.  By a suitable strategy for varying the time headways of ACC
systems as a function of the traffic situation, the unfavorable human
behavior can be partially compensated for. First simulations of such
ACC-systems show promising results \cite{Treiber-aut}.


\textbf{Acknowledgments:}
The authors would like to thank for partial support by the Volkswagen
AG within the BMBF project INVENT.





\end{document}